\documentstyle[12pt,psfig,epsf]{article}

\def\be{\begin{equation}}
\def\bea{\begin{eqnarray}}
\def\ee{\end{equation}}
\def\eea{\end{eqnarray}}
\def\to{\rightarrow}

\def\ra{{\rangle}}
\def\la{{\langle}}
\def\r{\right}
\def\l{\left}

\def\min{{\rm min}}

\def\a{\alpha}
\def\b{\beta}
\def\g{\gamma}
\def\lam{\lambda}

\def\d{{\rm d}}

\begin{document}

\title{Biodiversity in model ecosystems, I:
Coexistence conditions for competing species.}

\author{Ugo Bastolla$^1$, Michael L\"assig$^2$, Susanna C. Manrubia$^1$,\\
and Angelo Valleriani$^3$}

\maketitle

\begin{center}
{\it $^1$ Centro de Astrobiolog\'{\i}a, INTA-CSIC, Ctra. de Ajalvir 
km. 4, 28850 Torrej\'on de Ardoz, Madrid, Spain.
$^2$ Institut f\"ur theoretische Physik, Universit\"at zu K\"oln, Z\"ulpicher 
Strasse 77, 50937 K\"oln, Germany. 
$^3$ Max Planck Institute of Colloids and Interfaces, 14424 Potsdam, Germany.}
\end{center}

\begin{abstract}
This is the first of two papers where we discuss the limits imposed by
competition to the biodiversity of species communities.
In this first paper we study the coexistence of competing species at 
the fixed point of population dynamic equations.
For many simple models, this imposes a limit on the width of the
productivity distribution, which is more 
severe the more diverse the ecosystem is (Chesson, 1994).
Here we review and generalize this analysis, beyond the ``mean-field''-like
approximation of the competition matrix used in previous works, and extend
it to structured food webs.
%
In all cases analysed, we obtain qualitatively similar relations between
biodiversity and competition: the narrower the productivity distribution
is, the more species can stably coexist. We discuss how this result,
considered together with environmental fluctuations, limits the maximal
biodiversity that a trophic level can host. 
\end{abstract}


\section{Introduction}

One of the most striking characteristics of living systems is their amazing
diversity. Theoretical ecologists have devoted much of their effort to
explain why Nature is diverse and to identify the mechanisms that enhance
or limit species coexistence. While the observations show that rich and
diverse ecosystems are the rule and not the exception, stable, highly
diverse systems rarely arise in mathematical models.

In the early decades of theoretical ecology, emphasis has been placed in the
{\it principle of competitive exclusion}. Stated qualitatively, this principle
asserts that two or more species occupying the same ecological niche cannot 
stably coexist in the same ecosystem.
The first attempt to transform this principle into a mathematical theorem
is due to Volterra (1928). 
Generalizations of Volterra's
theorem to an $S$ species ecosystem have been provided by several
authors (MacArthur and Levins, 1964; Rescigno and Richardson, 1965; Levin, 
1970). It has been shown that $S$ species cannot coexist
at a fixed density if they are limited by less than $S$ independent
resources.

This approach, however, presents two kinds of shortcomings. The first 
one is of mathematical nature. It has been shown  that the
theorem of competitive exclusion in the previous form does not hold 
if some 
conditions are relaxed. Most importantly, it no longer applies if the growth
rates depend non-linearly on the resources and if coexistence
at a fixed point is replaced by the more general condition of persistence
(Armstrong and McGehee, 1980; Koch, 1974; Kaplan and Yorke, 1977).
Such non-linear behaviour 
is to be expected in natural ecosystems.  
Similarly, the theorem of competitive exclusion is violated when
age structured populations are considered,
when resource is not uniform in quality (May, 1974; Diamond, 1978) or when
physical space is considered (Sol\'e {\it et al.}, 1992). 
The prolonged coexistence of a large number of competitive plankton species
has been justified through barriers to mixing of species in an otherwise
homogeneous environment (Bracco {\it et al}., 2000).

The second, more important difficulty is related to field observations,
which can hardly ever claim to give a full account of all resources 
and the functional dependence of growth rates on them.
Hence, the theorem of competitive exclusion has little actual 
predictive value. The simultaneous presence of similar species can 
always be ascribed to yet unknown resources or to unknown functional 
forms of growth rates. 


Field research has shown that coexistence of competing species is
far from rare in real ecosystems. As a result, the conditions favoring 
coexistence are receiving increasing attention from the ecological community
(McCann {\it et al.}, 1998; Chesson, 2000; Kokkoris {\it et al.}, 2002;
Roberts and Stone, 2004).
These theoretical analysis and empirical observations suggest that the 
competitive exclusion principle could be replaced by a `coexistence 
principle' (Boer, 1986). 

In the present suite of two papers, we analyze species coexistence for a
class of models where competitive exclusion does not hold, since species
growth is limited both by biotic resources (explicitly represented into the
model) and by other limiting factors (implicitly considered), modeled as
self-damping terms in the population dynamics equations.
In the first paper we consider coexistence at the fixed point of population
dynamics, and show how the combination of competition and environmental
fluctuations limit the maximum amount of biodiversity that a trophic level
can host. In the second paper we consider models of species assembly, kept
away from the fixed point of population dynamics through the continuous
arrival of new species due to immigration or speciation events. These models
can reproduce species-area laws in good agreement with field observations.
The coexistence condition presented in this paper can be generalized to that
situation as well. We define therefore an effective model of biodiversity
across a food web, based on an approximation to the population dynamics
equations and on the condition of maximum biodiversity derived in this
first paper. The effective model predicts that biodiversity as a function
of the trophic level has a maximum at an intermediate level.

The present paper is organized as follows. After introducing the population
dynamics equations that we use, we discuss in section~3 coexistence in a
single trophic layer, in the framework of a
mean-field approximation of the competition matrix.
The coexistence condition that we derive imposing that all equilibrium
densities are positive is equivalent to the one demonstrated by Chesson
through the criterion of invasibility (Chesson, 1994; 2000).
Several different models yield the same coexistence condition,
be competition represented either explicitly in the population dynamics
equations or implicitly, through the effect that other species have on
resources. This condition states that
species coexistence depends crucially on the
{\em distribution of rescaled net productivity}, i.e. productivity not
taking into account the competition load and rescaled through the
carrying capacity:
The productivity distribution has to get narrower
for allowing the coexistence of a larger number of species.
This result is also in agreement with theoretical
studies stating that many species can coexist provided they are 
similar enough (Kokkoris {\it et al.}, 2002). 

The coexistence condition is then generalized to generic competition
matrices, beyond the mean-field approximation. 
We show that the angle formed by the principal eigenvector of the
competition matrix, which we name the competition load, and the rescaled
productivity must be narrow to permit species coexistence. This result
allows to generalize the coexistence condition from one trophic layer to
structured food webs, as discussed in Section~5.


The coexistence condition alone does not impose a limit on biodiversity
if the productivity distribution can be arbitrarily narrow. 
However, in natural ecosystems this distribution has a finite width
due to unavoidable environmental fluctuations on time scales much shorter
than those of population dynamics. This limits the maximal biodiversity
the system can host, and produces a typical shape of biodiversity versus
trophic level (L\"assig {\it et al.}, 2001), as we discuss in the companion
paper. 

\section{General framework}



We study here models of multispecies communities.
A key ingredient in the models is biodiversity,
meaning the number of reproductively separated populations in
the environment. Biodiversity arises from a balance between the assembly
process involving speciation and immigration and the extinction process
driven by population dynamics, in a spirit similar to MacArthur and Wilson's
theory of island biogeography (MacArthur and Wilson, 1967).
Population dynamics is represented through generalized Lotka-Volterra
equations of the form

\be
\frac{1}{N_i} \frac{\d N_i}{\d t} =
\sum_{j \ne i} g_{ij} ({\bf N}) - \sum_{j=1}^S\b_{ij} N_j(t) - \a_i \,
\,
\label{popdyn}
\ee
where the community is formed by $S$ interacting species,
$N_i$ represents the biomass density of species $i$, and
${\bf N}=\{N_1, N_2, \dots, N_S\}$.

The function $g_{ij} ({\bf N})$ models prey-predator relationships, and is
called {\it Predator Functional Response} (PFR).
It is zero for pairs of species not connected by a predator-prey relationship.
If species $i$ is a predator and species $j$ is its prey, $g_{ij}$ is
positive and represents the rate of prey consumption per unit of predator
biomass. The sign is negative if species $i$ is a prey and $j$ is its
predator. We assume in this case $g_{ij}({\bf N}) N_i=
-g_{ji} ({\bf N}) N_j/\eta$. The term $\eta\leq 1$ is the efficiency of
conversion of prey biomass into predator biomass. In the present paper
we set $\eta=1$, in order to simplify formulas. In the companion paper
the factor $\eta$ will play a more important role, and it will be
explicitly indicated.

Several different mathematical forms of the
PFR have been proposed and discussed in the biological literature. We study
here two cases: (i) prey dependent PFR linear in prey
density, i.e. $g_{ij}({\bf N})=\g_{ij} N_j$; (ii) ratio dependent PFR, i.e.
$g_{ij}$ is a function of the ratio between prey and predator density
(Arditi and Ginzburg, 1989).

The matrix $\b_{ij}$ models competition as a linear reduction in growth
rates\footnote{This linear model can be 
understood as the linearized version, around an equilibrium point, of a more 
general model, as far as equilibrium properties are captured in the 
linearized representation.}.
All its elements are non-negative, and the diagonal elements $\b_{ii}$ are
different from zero.  These intraspecific competition terms play an essential
role in allowing coexistence.
Since competition for prey is already represented through the
terms $g_{ij}$ and the dynamics of prey species, the terms $\b_{ij}$ stand
for competition for resources not explicitly
included in the model. Such terms naturally arise from ``integrating out''
some trophic links in a community.
The term $\a_i>0$ accounts both for the death rate and
for the energy consumption necessary for the activity of species $i$.

Equations (\ref{popdyn}) are complemented with the
condition that species below a critical density $N_c$ get extinct.
This condition takes into account that species are composed of discrete
entities and also mimics the effect of demographic stochasticity.

We represent explicitly a single external resource, interpreted as abiotic
and considered as an additional ``species'' $N_0(t)$ (Caldarelli {\it et al.},
1998; Bastolla {\it et al.}, 2001). Its dynamics 
is chosen in such a way that species feeding on it feel an
indirect competition effect. The qualitative behavior of the model does not
depend on the detailed dynamics of the abiotic resource, as far as
competition in the first trophic level is properly represented.
The external resource introduces a new scale of density $R$. The
dimensionless quantity $R/N_c\gg 1$ is an important control parameter
in the system.

\section{One-layer competition and productivity distribution}
\label{sec_ine}

In this section, we reformulate results showing that the coexistence
of several competing species tends to equalize their net productivity: the
more coexisting species, the more similar their productivity should be 
(Chesson, 1994; 2000; L\"assig {\it et al.}, 2001; Kokkoris {\it et al.},
2002).
This is done here imposing that the $S$ competing species coexist at a
fixed point of population dynamics, with all densities positive and
larger than the threshold for extinction, $N_c$.
We adopt a simple mean-field approximation of the competition matrix,
which will be relaxed in next section.

The condition derived in this way is qualitatively
equivalent to a condition derived through the more general requirement of
invasibility (Armstrong and McGehee, 1980; Chesson, 1994; 2000).
Moreover, modelling competition through explicit interaction terms or
implicitly, through the dynamics of the common resources, leads to the
same condition.

\subsection{Direct competition: mean-field approximation}
\label{caseC}

Our study of competition starts from the simplest model

\be
{1\over N_i}{\d N_i\over\d t}=
P_i-\sum_{j}\b_{ij} N_j\,\label{EqC} \, .
\ee
The quantity $P_i$ is assumed to be independent of species density, and
represents the intrinsic growth rate
of species $i$ in the absence of competition.

We assume that the competition matrix $\b_{ij}$
is symmetric and all its elements are non-negative. We further assume
for convenience that the matrix is positive definite. It can be
proven that these hypothesis hold if the competition terms arise in an
effective way through the dynamics of underlying resources. 
The stability properties of the fixed point $N_i^*=\sum_k(\b)^{-1}_{ik}P_k$
are governed by the community matrix (May, 1974), which in the present case
is $A_{ij}=-N_i^* \b_{ij}$. For positive definite competition matrices,
positivity of all the $N_i$ implies that the community matrix is negative
definite, thus the system is locally stable. Furthermore, under these
conditions, it is possible to construct a Ljapunov function (MacArthur and
Levins, 1964; May, 1974), which guarantees that the equilibrium point is
globally stable as well. Thus, we can ignore coexistence along periodic as
well as chaotic orbits.

%

We parameterize the competition matrix as

\be 
\label{betarho}
\b_{ij}= \sqrt{\b_{ii}\b_{jj}} \rho_{ij}\, ,
\ee
where $\rho_{ij}\in [0,1]$ is a dimensionless quantity that we call
{\it ecological overlap} (or niche overlap) and describes the similarity
in the use of resources between species $i$ and $j$. One has clearly
$\rho_{ii}=1$.

The parameters $\b_{ii}$ can be absorbed introducing the rescaled variables
$n_i=(\b_{ii})^{1/2} N_i$, $p_i=(\b_{ii})^{-1/2} P_i$.
The variables $p_i^2$ have dimensions of biomass per area
per time, the same dimensions of productivity. By analogy, and as a
shortening, we call the $p_i$'s productivities instead of rescaled growt
rates, but they should not be mistaken for
productivities measured in field studies.
In terms of the new variables, the fixed point equations have the form

\be p_i=\sum_j \rho_{ij}n_j\, .\label{fix1} \ee

We start by considering a mean field approximation of the competition
matrix, addressing the general case in the next section. The mean field
approximation consists in assuming that all non-diagonal elements are
equal: $\rho_{ii}\equiv 1$, $\rho_{ij}\equiv \rho_0$ for $i\neq j$,
\footnote{In the context of May and MacArthur's theory of competition,
this hypothesis is realized when the niche space has many 
dimensions (May and MacArthur, 1972).}
such that the solution of the fixed point equations reads

\be \label{comp-1}
n_i^*={1\over (1-\rho_0)}
\left(p_i- {\langle p\rangle\over 1+(1-\rho_0)/\rho_0 S}\right)\, ,
\ee
where $\langle \:\rangle$ indicates the average over the $S$ species in
the community. All equilibrium densities are positive and above the threshold
$n_c$ if and only if

\be \label{R1}
{\langle p\rangle  -p_i \over\langle p\rangle}
\leq {1- n_c/\langle n\rangle \over 1+S\rho_0/(1-\rho_0)}\, ,
\ee
For $n_c=0$, this result is equivalent to the condition derived by Chesson
(1994) imposing invasibility of the system.

The above condition only affects explicitly the $p_i$'s smaller than the
average, but it is easy to see that it implies a condition on the variance
of the productivity distribution. In fact, multiplying both sides by
the quantity $\l(1-(p_i-\langle p\rangle)/\langle p\rangle \r)$, which
is always positive if the variance of the distribution is small enough, 
and averaging over all species, we find

\be \label{R1-2}
{\langle p^2\rangle  - \langle p\rangle^2 \over\langle p\rangle^2}
\leq {(1-\rho_0)\over S\rho_0+(1-\rho_0)}
\left(1- \frac{n_c}{\langle n\rangle}\right) \, .
\ee

In other words, the coexistence condition requires that all $p_i$'s
are very close to the average value when the number of species $S$ is
large. The maximal negative difference from the average is of order
$1/S$ and the standard deviation is of order $1/\sqrt{S}$.
Notice that, if the mean overlap $\rho_0$ equals unity, all $p_i$
must be identical. This result corresponds to the formulation of the
theorem of  competitive exclusion in this framework.

For a resource rich system where the average reduced density
$ \langle n \rangle$ is well above the extinction threshold, the quantity
$S_0=(1-\rho_0)/\rho_0$ defines an intrinsic scale of biodiversity at which
the productivity distribution gets pretty narrow. 
For much larger number of species $S$, the average reduced density
$ \langle n \rangle = \langle p \rangle / \left(1+(S-1)\rho_0\right)
\propto \langle p \rangle /(1+S/S_0)$ is close to the extinction threshold,
thus the r.h.s. of Eq.~(\ref{R1}) becomes very small and the condition of
coexistence becomes rather stringent. Therefore, the effect of positive
$n_c$ on biodiversity is only important for very diverse ecosystems.


\subsection{Resource mediated competition, prey dependent PFR}
\label{caseA}

We now consider $S$ basal species feeding on a single external
resource $N_0$ with a linear, prey dependent 
PFR: $g_{ij}(\{N_k\}) = \g_{j} N_j$.
Competition is induced through the dynamics of the common resource
$N_0$ and through the limiting factors not explicitly considered in the model.
The population dynamics equations are 

\bea
{1\over N_i} {\d N_i\over \d t}&=&
\g_{i} N_0-\sum_j \b_{ij}N_j-\a_i \nonumber \\
{1\over N_0} {\d N_0\over \d t}&=& R\beta_0 
- \beta_0 N_0 -\sum_{i=1}^S \g_{i} N_i\, .
\eea
Different equations for $N_0$ give qualitatively similar results, as far
as $N_0$ is consumed by all the competing species. 
We consider the fixed point equations and substitute for $N_0$, finding

\be
\label{c2}
p_i -\sum_j\left({\g_i^\prime\g_j^\prime\over \b_0}+\rho_{ij}\right)n_j
=0\, ,
\ee
where we have defined the rescaled variables
$\g_i^\prime=\g_i/\sqrt{\beta_{ii}}$, $\a_i^\prime=\a_i/\sqrt{\beta_{ii}}$,
$p_i=\gamma_i^\prime R-\a_i^\prime$ and $n_i=N_i\sqrt{\beta_{ii}}$. 
The fixed point equations are thus equivalent to the equations obtained
by considering direct competition, with an effective competition matrix that
is symmetric and positive definite if $\b_{ij}$ is such.

We now adopt the mean-field approximation of the previous section,
$\rho_{ii}\equiv 1$, $\rho_{ij}=\rho_0<1$ ($i\neq j$).
The approximation is applied only to the matrix $\rho_{ij}$, since the
other part of the competition matrix depends on the productivity vector,
nevertheless the result turns out to be equivalent to the previous one.
Substituting $\g_i^\prime=(p_i+\a^\prime)/R$ in Eq.~(\ref{c2}), we get


\be
(1-\rho_0)n_i=
p_i(1-S\la n\ra a)-\a_i S\la n\ra a -\rho_0 S\la n\ra\, ,
\ee
where $a=\langle \g^\prime n\rangle/\la n\ra \b_0 R$.
We then neglect the dependence of $a$ on $\la n\ra$ and solve for this
variable. Rearranging the various terms, we find again that the
coexistence of $S$ species is only possible if the minimal productivity
differs from the average at most in an amount $(1-\rho)/\rho S$:

\be
\frac{p_i-\la p\ra}{\la p\ra}-
\frac{a\la \a^\prime\ra}{a\la \a^\prime\ra+\rho_0+(1-\rho_0)/S}
\frac{\a_i^\prime-\la \a^\prime\ra}{\la \a^\prime\ra}
\leq
{1- n_c/\langle n\rangle \over 1+S\rho/(1-\rho)}\, ,
\label{R1-ind}
\ee
where the effective competition overlap $\rho$ is defined as

\be
\rho={\rho_0 + a \la \a^\prime\ra \over 1+a \la \a^\prime\ra} \leq 1
\label{coe}
\ee

This coincides with Eq.~(\ref{R1}) in the limit
$ a \la \a^\prime\ra=
\la \a^\prime\ra \la \g^\prime n\rangle/\la n\ra \b_0 R\rightarrow 0$,
in which resources are not limiting.
In the opposite limit the condition on minus the energy consumption rate
$-\a^\prime_i$ becomes more demanding than the condition on the $p_i$'s,
and $\rho$ tends to one.
Notice that the competition overlap $\rho$ is different from zero even if
the explicit competition overlap $\rho_0$ vanishes.

\subsection{Resource mediated competition, ratio dependent PFR}
\label{caseB}

We now assume that the PFR
depends on the ratio between the density of the prey and the density
of its predators (Arditi and Ginzburg, 1989) as

\be g_{i0}({\bf N})= 
{b_0 c_{i} N_0\over b_0 N_0+\sum_{p\in P(0)}c_{p}N_p}\, . \label{2}
\ee
(see also Schreiber and Guti\'errez, 1998, Sol\'e {\it et al.}, 2000, 
Drossel {\it et al.}, 2001).
In this expression species ``$0$'' is the prey and
the sum in the denominator runs over its predators, represented as the set
$P(0)$. In the case of just one predator, Eq.~(\ref{2}) can be seen
as a Holling type III PFR (Holling, 1959),
where the prey density at which the functional response
saturates is proportional to the predator's density. Notice that, unlike
prey dependent PFR, the ratio dependent PFR does not
contain any externally specified biomass scale. For this reason, it is in
some sense simpler than prey dependent PFR (Bastolla {\it et al.}, 2001).
We do not aim at discussing which form of the PFR
is most suitable to explain observational data (Abrams and Ginzburg, 2000),
but just use it as a
second example of population dynamics where, aside of ratio dependence,
competition for preys is explicitly represented through the sum
over predators in the denominator. 
Since competition is now explicitly represented, we don't need an additional
equation for the external resource $N_0$ and we assume that it renews
rapidly enough such that its density remains constant. 
The dynamical equations have now the form

\bea
{1\over N_i}{\d N_i\over \d t} &=&
{b c_i N_0\over b N_0 +\sum_j c_j N_j}-\a_i -
\sum_j \beta_{ij} N_j\, , \nonumber \\
N_0(t) & \equiv & R \, .
\eea
To simplify formulas, we use the rescaled variables
$c_i^\prime=c_i/\sqrt{\b_{ii}}$, $n_i=N_i\sqrt{\b_{ii}}$ and apply
the mean field approximation to the matrix $\b_{ij}$. The solution of the
fixed point equations is then

\be
n_i^*=\langle n\rangle+{1\over (1-\rho_0)}
{c_i^\prime-\langle c^\prime\rangle\over 1+S a\la n\ra} \, ,
\ee
where now $a=\la c^\prime n\ra/\la n\ra bR$. To simplify formulas,
we consider the case of a productive system where the variation in
the $\a_i$s can be neglected.
The condition that all species are viable yields
\be \label{R3}
{\langle c^\prime\rangle - c^\prime_i \over \langle c^\prime\rangle}
\leq {1-n_c/\langle n\rangle \over 1+S\rho/(1-\rho)}\: ,
\ee
where
\be
\rho={\rho_0+\la \a^\prime\ra/\b S\la n\ra
\over 1+\la \a^\prime\ra/\b S\la n\ra}
\ee
is the effective competition coefficient and
$S\langle n\rangle$ can be obtained solving a second order equation, whose
root remains finite in the limit $S \to \infty$.

Equations (\ref{R1}), (\ref{R1-ind}), and (\ref{R3}) are equivalent
coexistence conditions for three different population dynamics models.
They all show the quantitative dependence between the distribution of
productivities and biodiversity.
Qualitatively identical coexistence conditions were obtained through the
invasibility criteria, i.e. imposing that an invading species (with very
low density) has a positive growth rate (Chesson, 2000). 

\section{One layer competition without mean-field}
\label{eigen}
In this section, we analyze the coexistence condition without relying
on the mean field approximation of the competition matrix.
Our starting point are the fixed point equations (\ref{fix1}), where
rescaled variables are used. To simplify the presentation, most calculations
are reported in Appendix~A. Here we only show and discuss the final results.

We will use the spectral representation of the matrix $\rho_{ij}$,
which we assume to be symmetric and positive definite. Therefore,
all of its eigenvalues $\lambda_\a$ are real and positive, and its $S$
eigenvectors ${\bf u}^\a$ form an orthonormal system.
In graph theory (Bollob\'as, 1998) the matrix $\rho_{ij}$ is called the
adjacency matrix. Its maximal eigenvalue is defined by the property
$\lam_1 \sum_i v_i^2 \geq \sum_{ij}\rho_{ij}v_i v_j$ for every vector
$v$. The equality holds if and only if $v$ is proportional to the
principal eigenvector ${\bf u}^1$. It holds that
$\lam_1 \geq \sum_{ij}\rho_{ij}/S$, so that the main eigenvalue is
expected to be proportional to the number of species $S$.
Moreover, all components of the principal eigenvector either have the same
sign (we can choose it arbitrarily to be the positive sign) or are zero.
We will assume that the graph cannot be separated in disconnected
components. If this is not the case, the analysis can be applied separately
to each disconnected component. This hypothesis implies that all components
of the principal eigenvector are strictly positive: $u^1_i>0$.
We call the principal eigenvector of the competition matrix the
{\it competition load}.

It is useful to define the average of the $S-1$ eigenvalues excluding the
principal one, which we will denote with the symbol $1-\rho_0$:
\be 1-\rho_0 \label{rho-eigen}
\equiv \frac{1}{S-1}\sum_{\a=2}^S\lam_\a\, . 
\ee

It follows from the condition on the trace of the matrix $\rho_{ij}$ that
$\sum_\a\lam_\a=S$, so the principal eigenvalue $\lam_1$ can be
expressed as

\be
\lam_1=S\rho_0+\left(1-\rho_0\right)\: .
\ee

We now use for simplicity the notation $v^\a\equiv \sum_j v_j u^\a_j$ for
the projection of a vector ${\bf v}$ along the $\a$-th eigenvector,
${\bf u}^\a$.
We show in Appendix~A that the coexistence condition
(\ref{R1}) can be written, for a general competition matrix, as

\be\label{R1-bis}
\frac{\sum_i p_i^2-(p^1)^2}{(p^1)^2}  \leq
\frac{\lam_2}{S\rho_0+\left(1-\rho_0\right)}
\left(1- \frac{n_c}{\langle n\rangle} \right)\, ,
\ee
where $\lam_2$ is the second largest eigenvalue, which is of order one.

If the matrix $\rho_{ij}$ has the mean field form, Eq.~(\ref{R1-bis})
is identical to Eq.~(\ref{R1-2}). In fact, in the mean field case, it
holds that $\lam_\a=(1-\rho_0)$ for $\a\geq 2$.
Moreover, in this symmetric system the principal eigenvector is uniform,
$u^1_i\equiv 1/\sqrt{S}$, so that $p^1=\sqrt{S}\langle p\rangle$, yielding
Eq.~(\ref{R1-2}).

Eq.~(\ref{R1-bis}) tells us that the
productivity vector, $p_i$, has to be almost parallel to the competition
vector $u^1_i$ in order to allow coexistence of all species:
Species with larger competition component $u^1_i$ need a comparatively
larger productivity to survive.
The angle between the productivity vector and the competition vector has to
become narrower as competition becomes stronger, until only a perfect
coincidence guarantees the survival of all species. The strength of
competition increases with the overlap parameter $1-\lam_2$, with the ratio
$n_c/\langle n\rangle$, and with the largest eigenvalue
$\lam_1\approx S\rho_0$. Therefore, competition becomes more severe as
the number of species increases, unless the competition matrix is sparse.
Since the resources on which the $S$ species feed are finite, we
expect on ecological grounds that the strength of the competition 
increases as more species are packed in the ecosystem. This can
be shown in some mechanistic models of competition, in which
the matrix $\rho_{ij}$ is derived from the underlying dynamics of
the resources.

\section{Competitive coexistence in a food web}

We now turn to the more general case of an $L-$levels food web.
After integrating out the upper and lower levels,
the effective equation for the fixed point density of species at a given
level has the familiar form of a competition equation, and displays the
same qualitative behaviour discussed for the case of the single level.
 
As in our previous work (L\"assig {\it et al.}, 2001), we assume a
hierarchical trophic
organization, so that species at level $l$ feed only on species at level
$l-1$ and compete only with species at their same level. 
Species at level $L$ are top predators. Level zero can be interpreted either
as basal species or as abiotic resources described by an effective equation.
The dynamical equations, using linear, prey dependent PFR, read

\bea
{1\over N_i^{(l)}} {\d N_i^{(l)}\over \d t} & = &
\sum_j\g_{ij}^{(l)}N_j^{(l-1)}-\a_i^{(l)}  \label{web_eq} \\
& - & \sum_j \b_{ij}^{(l)}N_j^{(l)}-\sum_j\g_{ij}^{(l+1)}N_j^{(l+1)}\: ,
\nonumber
\eea
where the superindex stands for the level where the species belongs.
The matrix $\g_{ij}^{(l)}$ represents predation from level $l$ to level
$l-1$. It vanishes identically for $l=0$ and $l=L+1$. The parameters 
$\a_i^{(l)}$ stand for energy consumption and death rates of species at
level $l$ for $l \ge 1$, whereas $-\a_i^{(0)}$ is the fictitious growth
rate of the resource at level zero. The direct competition for species 
at level $l$ is represented through the matrix $\b_{ij}^{(l)}$. As above,
we assume that it is symmetric and positive definite. 

We now consider the fixed point equations. For every level $l$, we solve for
the densitites of species at different levels $l'\neq l$ and substitute,
thus getting fixed point equations for species at level $l$ with the form of
effective competition equations with a symmetric competition matrix:

\be
\sum_j C_{ij}^{(l)}N_j^{(l)}=P_i^{(l)} \: .
\label{web_comp}
\ee
Three different integration schemes yield different equations of the
same form (\ref{web_comp}). In the scheme described in Appendix~B,
densities at levels $l'<l$ are recursively solved starting from level zero,
and densities at levels $l'>l$ are recursively solved starting from the
maximum level $L$.
Levels above and below $l$ (predators and preys) contribute both to the
competition matrix and to the productivity vector. Therefore, as it is
known, competition can be induced not only through common preys, but
also through common predators.

In a second possible scheme, levels above $l$ only contibute to $P^{(l)}$
as an effective energy consumption term, whereas levels below $l$
contribute both to the competition matrix and to the productivity
vector as a growth term. Finally, in the third scheme, levels above $l$
contribute both to the effective competition and to the productivity vector,
whereas species at levels below $l$ only contribute to the productivity.
In all three cases, the effective equations for species at level $l$ have
the form (\ref{EqC}) of effective competition equations, where the
competition matrix and the productivity vector depend
on the properties of species at the other levels.

The result of the previous section implies that also for a structured food
web the effective productivity vector at level $l$ must form a narrow
angle with the principal eigenvector of the effective competition matrix.

\section{Environmental variability and biodiversity}
\label{sec:delta}


The coexistence conditions derived above yield a natural scale for assessing
the degree of biodiversity of an ecosystem\footnote{
This scale is relevant in ecosystems
where $\langle N\rangle \gg N_c$, otherwise biodiversity is controlled by the
threshold density. We are assuming here that our ecosystem is large enough
so that the condition $\langle N\rangle \gg N_c$ holds where the number of
species is of the order $(1-\rho)/\rho$ at which competition effects are
important.}.
Ecosystems with $S\ll (1-\rho)/\rho$ are
{\it loosely packed}. They impose mild conditions on the productivity
distribution, and can easily incorporate new species. Ecosystems with
$S\gg (1-\rho)/\rho$ are {\it tightly packed}. There,
the productivity distribution is subject to strict constrains and
incorporation of new species is very difficult. Which ecological
mechanisms distinguish between the two kind of situations? On the one hand,
as we will discuss in a next paper, the immigration and speciation rates
play a key role in smoothing the effects of condition (\ref{R1}) and
allowing a large number of species into the system. On the other hand,
an important role in determining biodiversity is played by environmental
fluctuations (Hutchinson, 1961; Chesson, 2003a). 


The growth rates considered in the previous models are subject in real
systems to environmental fluctuations, such as
for instance variation in rain levels, temperature and daily light, or
forest fires for plant communities, natural obstacles to the movement
of animal species, or fluctuations in the number of interacting
populations also affected by the environmental noise. Some of these
variables fluctuate over time scales much shorter than the characteristic
time scales of population dynamics. We thus interpret the growth rates
$p_i$'s as time averaged quantities with superimposed fast fluctuations
due to environmental variability. These preclude that the productivities
take identical values, even for populations with identical ecological
properties. 


To take into account these fast environmental fluctuations, we assume that
they tend to broaden the productivity distribution so that there is a
minimum width given by the equation

\be
{\langle p\rangle - p_{\rm min} \over \la p\ra}
\geq \Delta\left(1-{1\over S}\right) \label{fluct}
\ee
The factor $1-1/S$ ensures that the condition (\ref{fluct}) is satisfied
for $S=1$, when $P_\min$ and $\la P\ra$ coincide.
The term $0<\Delta \leq 1$ is an effective measure of environmental
variability. For $\Delta=0$ the condition (\ref{fluct}) is always satisfied,
while for $\Delta=1$ it imposes a very small $P_\min$ (recall the $P$'s are
positive quantities).
Values of $\Delta>1$ can not be realized in the large $S$ limit.

Combining Eq.~(\ref{fluct}) with the coexistence condition (\ref{R1}), we
get a limit to the maximal biodiversity that the ecosystem can host.
In fact, Eq.~(\ref{R1}) requires that the productivity distribution gets
narrower as biodiversity increases, but at some point the largest difference
tolerated becomes of the order of the minimal difference compatible with
the actual environmental variability. The maximal biodiversity is a function
of the competition overlap $\rho$, of the environmental variability $\Delta$,
and of the ratio between the average biomass and the threshold for
extinction, $n_c/\la n\ra$, and is given by the inequality

\be
{1-n_c/\la n\ra \over 1+S\rho/(1-\rho)}
\geq \Delta \left(1-{1\over S}\right)\: ,
\ee
leading to a second order inequality
%
that can be simplified for large $S$ as
\be
\label{Slargeenv}
S\leq 1+\left({1-\rho\over\rho}\right)
\left({1-\Delta-n_c/\la n\ra\over \Delta}\right)
\ee
For large competitive overlap $\rho$ close to unity or large variability
$\Delta \simeq 1$, only one species can survive in the long run.
If both the overlap and the variability are small, on the other hand,
the maximal number of species can be rather large. Only in case of
small variability the ecosystem can be tightly packed.

\section{Discussion}

In this work, we have considered simple models of competition,
either represented as explicit terms in the population dynamics equations
or effectively introduced through the dynamics of common prey and predator 
species. We have focused our attention on the fixed points of population
dynamics, imposing the condition that all species at equilibrium have positive
densities. For the simplest model of competition, this condition ensures
that the fixed point is both locally and globally stable, but for more
complex situations this represents an oversimplification.
We will consider in the companion paper systems far from equilibrium, and
argue that a suitable modification of the coexistence condition derived
at the fixed point remains generally valid.

We first analyzed coexistence at the fixed point, adopting a mean field
approximation of the competition matrix. This approximation is expected to
hold when there is a main nutrient on which all species are dependent, or
when the niche space has many dimensions.

Within the mean-field approximation, coexistence of $S$ species competing
with each other implies a condition on the distribution of their rescaled
growth rates, $p_i=P_i/\sqrt{\beta_{ii}}$, where $\beta_{ii}$ is the
inverse of the carrying capacity: The width of the distribution has to shrink
with increasing number of species $S$, competition overlap $\rho$ or ratio
$N_c/\la N\ra$ of the minimal viable density to the average density.
Notice that, if the carrying capacities vary from one species to
another, for instance depending on body size, this conclusion does not
apply to growth rates prior to rescaling.

This coexistence condition is equivalent to the one obtained by
Chesson (1994; 2000) by imposing the requirement of invasibility,
which demonstrates its robustness.
We found formally identical coexistence conditions in models with explicit
competition terms and in models where competition is induced by the dynamics
of shared resources, within a single trophic layer and
within a whole structured food web.

The mean field approximation is not appropriate to describe more
complex competition matrices, for instance when there is a continuum of
resources distributed along one dimension (May and MacArthur, 1972), or
when species can be grouped according to their ecological similarity.
We have therefore generalized the coexistence condition beyond mean-field
competition matrices. In the general case, an important role is played
by the principal eigenvector of the competition matrix, which we named the
competition load. The angle between the vector of competition loads and
the vector of rescaled growth rates has to be narrower for stronger
competition, i.e. for increasing average competition overlap and number of
species. This generalizes the condition on the variance of the distribution
of rescaled growth rates to the case of a generic matrix.

The coexistence condition Eq.~(\ref{R1-2}), stating that the variance of
the rescaled growth rates must decrease as $1/S$ due to competition, is 
reminiscent of May's theorem, according to which the coexistence of $S$
species randomly interacting requires that the variance of their interactions
vanishes as $1/S$ (May, 1974). In fact, May's theorem was derived
imposing local stability of the fixed point of population dynamics and, for
the simplest models of competition, the condition that all densities are
positive at the fixed point also implies local stability.
The condition for the coexistence of species whose rescaled growth rate lies
below the average is stricter: they must differ from the average at most by
an amount of order $1/S$, which is smaller than that imposed by the
condition on the variance. This appears more demanding than the condition
in May's theorem, probably because here all species are in competition with
each other instead of being randomly interacting.

The coexistence condition that we described does not limit biodiversity, as
long as the angle between rescaled growth rates and competition loads can be
made arbitrarily small. However, environmental fluctuations, due to
biological or abiotic processes with time scales shorter than those of
population dynamics, necessarily limit the possibility to fine tune
ecological parameters in order to accomodate a larger biodiversity.
We include environmental variations in our competition model in an effective
way, as a force broadening the distribution of rescaled growth rates.
Considering this new ingredient sets a limit on the maximal biodiversity
that a system of competing species can host.
In the companion paper, we will combine this limitation to ``horizontal''
biodiversity imposed by competition, with either dissipation of energy or
growth of perturbations in the ``vertical ''direction along the food chain.

Environmental changes characterized by longer time scales, as seasonal
changes or the storage effect, have not been included in our analysis,
although they certainly affect coexistence conditions. Interestingly, in
some cases the adaptative responses of species to these environmental
fluctuations with time scale comparable to that of population dynamics are
predicted to enhance species coexistence (Chesson, 2003a; 2003b). This
constrasts with the prediction presented here that a fast fluctuating
environment would set a limit on biodiversity. It would be therefore quite
desirable to set up a general theory of how environmental noise on
different time scales modulates biodiversity.

\section*{Acknowledgements}
UB, ML and SCM acknowledge hospitality and support by the Max Planck
Institut for Colloids and Interfaces during part of this work.
UB was also supported by the I3P program of the Spanish CSIC, cofunded
by the European Social Fund. SCM benefits from a RyC fellowship of
MEC, Spain.

\section*{Appendix A: Coexistence condition for a generic matrix}
\label{app-eigen}

Our starting point is Eq.~(\ref{fix1}), $p_i=\sum_j \rho_{ij}n_j$,
and we want to demonstrate the coexistence condition (\ref{R1-bis})
using the spectral properties of the competition matrix $\rho_{ij}$.
For simplicity of notation we shall indicate with a superscript $\a$ the
component of a vector in the direction of the $\a$-th eigenvector of the
matrix $\rho_{ij}$,  $n^\a=\sum_j n_j u^\a_j$. $\lam_\a$ denotes the
corresponding eigenvalue.
Using this notation, Eq.~(\ref{fix1}) can be written in the eigenvector
basis as $p^\a=\lam_\a n^\a$, from where we get the equilibrium biomasses

\be n_i=\sum_{\a=1}^S \frac{p^\a u^\a_i}{\lam_\a} \geq n_c\, .
\ee
We now bring to the r.h.s. the contribution of the principal
eigenvector $u^1_i$ and multiply both sides times the quantity
$\left[(p_i-p^1u^1_i)/p^1-u^1_i\right]$, which, as we will demonstrate
at the end, is always negative if system size is large enough. We get

\be
\sum_{\a=2}^S \frac{p^\a u^\a_i}{\lam_\a}
\left(\frac{p_i-p^1u^1_i}{p^1u^1_i}-1\right)
\leq \left(n_c-\frac{p^1 u^1_i}{\lam_1}\right)
\left(\frac{p_i-p^1u^1_i}{p^1u^1_i}-1\right)\, .
\ee

Summing over $i$, exploting the orthonormality of the eigenvectors,
and rearrange the factors, we obtain

\be
\sum_{\a=2}^S \frac{(p^\a)^2}{\lam_\a}
\leq \frac{(p^1)^2}{\lam_1}-
n_c \left(2p^1\langle u^1\rangle - \langle p\rangle\right)\, .
\ee

Here brackets denote average over the $S$ species. It holds
$\langle p\rangle= \sum_\a p^\a \langle u^\a\rangle$. Neglecting , just
for the sake of simplifying the final formula, the contribution of the
eigenvectors with $\a\geq 2$ to the mean productivity, which is small
since the components $p^\a$ and the mean eigenvectors $\langle u^\a\rangle$
are much smaller than the corresponding quantities for the principal
eigenvector, we get

\be
 \frac{\sum_{\a=2}^S(p^\a)^2/\lam_\a}{(p^1)^2}
\leq \frac{1}{\lam_1}\left(1-\frac{n_c}{\langle n\rangle}\right)\, .
\ee

Noticing that $\lam_2$ is the largest of all eigenvalues with $\a\geq 2$,
and that $\sum_{\a=2}^S(p^\a)^2=\sum_{i=1}^S p_i^2-(p^1)^2$, and
subsituting the expression for $\lam_1$, we finally get Eq.(\ref{R1-bis}):

\be
\frac{\sum_i p_i^2-(p^1)^2}{(p^1)^2}  \leq
\frac{\lam_2}{S\rho_0+\left(1-\rho_0\right)}
\left(1- \frac{n_c}{\langle n\rangle} \right)\, .
\ee
This expression demonstrates our previous statement that the quantity
$(p_i-p^1u^1_i)/p^1u^1$ is always smaller than one if system
size is large enough, which is the property that we have used for
deducing the above formulas.

\section*{Appendix B: Effective competition in a food web}
\label{app-web}

Starting from the food web equations (\ref{web_eq}), we want to obtain
effective fixed point equations for level $l$ in the form (\ref{web_comp}).
First, we write the fixed point equations in matrix notation:

\be
\left(\g^{(l)}\right){\bf N}^{(l-1)} - {\bf \a}^{(l)}-
\left(\b^{(l)}\right){\bf N}^{(l)}-\left(\g^{(l+1)}\right)^T
{\bf N}^{(l+1)}=0\, .
\ee

Bold face symbols represent column vectors, while the other symbols represent
matrices. An upper $T$ indicates tranposition of a matrix.
The boundary conditions are $\g^{(L+1)}=0$ (the top level $L$ has
no predators) and a formally identical equation with constant growth
rate for the renewable abiotic resources at level 0:

\be
\g^{(0)}R-\left(\g^{(1)}\right)^T{\bf N}^{(1)}=\b^{(0)}N^{(0)}\: .
\ee

We now focus on an intermediate level $l$.
Species below level $l$ can be integrated out solving for their equilibrium
densities starting from level 0 upwards. Species above level $l$ can be
integrated out similarly starting from the maximum level $L$. Both kinds
of species contribute to the effective competition matrix $C^{(l)}$
and effective productivity vector ${\bf P}^{(l)}$ of Eq.(\ref{web_comp}).
The recursive equations to obtain these quantities are

\bea
C^{(l)} & = &
\left(\g^{(l)}\right) \left(M^{(l-1)}\right)^{-1} \left(\g^{(l)}\right)^T
+ \b^{(l)} \nonumber \\
& + & \left(\g^{(l+1)}\right)^T \left(\tilde M^{(l+1)}\right)^{-1}
\left(\g^{(l+1)}\right) \\
{\bf P}^{(l)} & = &
\left(\g^{(l)}\right)  \left(M^{(l-1)}\right)^{-1} {\bf Q}^{(l-1)}
- {\bf \a}^{(l)}  \nonumber \\
& + & \left(\g^{(l+1)}\right)^T \left(\tilde M^{(l+1)}\right)^{-1} 
{\bf \tilde Q}^{(l+1)}
\eea
Quantities with the tilde are recursively obtained from the upper levels,
starting from level $L$:

\bea
\tilde M^{(l)}=
\b^{(l)}+ \left(\g^{(l+1)}\right)^T
\left(\tilde M^{(l+1)}\right)^{-1}
\left(\g^{(l+1)}\right)
\\
{\bf \tilde Q}^{(l)}=
 - {\bf \a}^{(l)} +
\left(\g^{(l+1)}\right)^T 
\left(\tilde M^{(l+1)}\right)^{-1}  {\bf \tilde Q}^{(l+1)} \, .
\eea
The boundary condition is $\left(\tilde M^{(L)}\right)=\b^{(L)}$,
${\bf \tilde Q}^{(L)}=-{\bf \a}^{(L)}$.
Quantities without the tilde are similarly obtained starting the 
recursion from level zero:

\bea
M^{(l)}=
\left(\g^{(l)}\right) \left(M^{(l-1)}\right)^{-1}
\left(\g^{(l)}\right)^T+ \b^{(l)} \\
{\bf Q}^{(l)}=
\left(\g^{(l)}\right)  \left(M^{(l-1)}\right)^{-1} {\bf Q}^{(l-1)} -
{\bf \a}^{(l)}
\eea
The boundary conditions are in this case $\left(M^{(0)}\right)=\b^{(0)}$
and ${\bf Q}^{(0)}=\g^{(0)}R$.

The effective competition matrix $C_{ij}^{(l)}$ is
positive defined if all of the $\b$ and $\g$ matrices are so.

\end{document}